# Title: One-dimensional moiré chains with partially-filled flat bands in two-dimensional twisted bilayer WSe$_2$

**Authors:** Ya-Ning Ren[1,2], Hui-Ying Ren[1,2], Kenji Watanabe[3], Takashi Taniguchi[4], and Lin He[1,2,]*

**Affiliations:**

[1]Center for Advanced Quantum Studies, Department of Physics, Beijing Normal University, Beijing, 100875, China

[2]Key Laboratory of Multiscale Spin Physics, Ministry of Education, Beijing, 100875, China

[3]Research Center for Functional Materials, National Institute for Materials Science, Tsukuba, Japan

[4]International Center for Materials Nanoarchitectonics, National Institute for Materials Science, Tsukuba, Japan

*Correspondence and requests for materials should be addressed to Lin He (e-mail: helin@bnu.edu.cn).

**Two-dimensional (2D) moiré systems based on twisted bilayer graphene and transition metal dichalcogenides provide a promising platform to investigate emergent phenomena driven by strong electron-electron interactions in partially-filled flat bands[1-11]. A natural question arises: is it possible to expand the 2D correlated moiré physics to one-dimensional (1D)? This requires selectively doping of 1D moiré chain embedded in the 2D moiré systems, which is an outstanding challenge in experiment and seems to be not within the grasp of today's technology. Therefore, an experimental demonstration of the 1D moiré chain with partially-filled flat bands remains absent. Here we show that we can introduce 1D boundaries, separating two regions with different twist angles, in twisted bilayer WSe$_2$ (tWSe$_2$) by using scanning tunneling microscopy (STM), and demonstrate that the flat bands of moiré sites along the 1D boundaries can be selectively filled. The charge and discharge states of**

**correlated moiré electrons in the 1D moiré chain can be directly imaged and manipulated by combining a back-gate voltage with the STM bias. Our results open the door for realizing new correlated electronic states of the 1D moiré chain in 2D systems.**

**Main Text:**

Twisted transition metal dichalcogenide (tTMD) materials with flat bands offer an unparalleled opportunity to explore exotic strongly correlated phases in two-dimensional (2D) moiré systems[1-8,12-18]. In the flat bands, the dispersion is strongly reduced, leading to enhanced electron-electron interactions for the emergence of correlated states. Usually, a back gate and/or a top gate are used to change the moiré global filling factor of the 2D system to realize the correlated states[1-18]. If one can control the filling factor of selective moiré sites locally rather than globally, it will be possible to expand the 2D correlated moiré physics to one dimensional (1D) or even fractional dimensional, which may lead to new emergent correlated electronic states that are absent in 2D moiré system. However, an experimental realization of such 1D or fractional dimensional moiré sites with partially-filled flat bands has not yet been achieved up to now. In this study, we report a facile method to controllably introduce 1D boundaries, separating two regions with different twist angles, in twisted bilayer $WSe_2$ (t$WSe_2$) by using scanning tunneling microscopy (STM) tip. Because of large local strain and lattice rotations, the doping along the 1D boundary differs quite from that of the two adjacent regions, which enables us to realize partially-filled flat bands only in the moiré sites along the 1D boundary. The strongly localized correlated electrons of the local moiré sites along the 1D boundary allow us to directly image the charge and discharge states by using STM[16,17]. Moreover, our experiment indicates that the charge states of correlated moiré electrons can be manipulated by combining a back-gate voltage with the STM bias.

By using micro-mechanical stacking technique[17,20], we fabricated a high-quality tWSe$_2$ device on a hexagonal boron nitride (hBN) substrate with a graphite back gate, as schematic shown in Fig. 1a (Methods). A monolayer graphene sheet is further covered on the tWSe$_2$ device to provide a conducting layer for collecting the tunnelling current from the STM tip in the STM measurements[21-23]. Our experiment indicates that the topmost graphene sheet almost does not affect electronic properties of the tTMD device, as reported in previous studies[13,16]. Figure 1b shows a large-scale STM topography image measured on the sample surface, exhibiting a quasi-1D superlattice with a period of about 100 nm, which is the superstructure of the hBN substrate possibly generated by the 1D crack of the hBN. A representative zoom-in STM measurement, as shown in Fig. 1c, reveals a moiré superlattice morphology with an approximate periodicity of 4.4 nm, formed by alternatingly rotating two WSe$_2$ layers with a twist angle $\theta \sim 4.3 \pm 0.1°$ (Extended Data Fig. 1). The 100-nm-period quasi-1D superlattice of the hBN substrate does not affect the moiré period of the tWSe$_2$, which is quite uniform in the region shown in Fig. 1b. Our scanning tunneling spectroscopy (STS) measurements shows a bandgap of about 2 eV in the tWSe$_2$. The moiré flat bands are observed at both the conduction band edge and valence band edge, as reported in previous studies[4,12] (Extended Data Fig. 2).

By applying a STM tip pulse, a nanoscale pit can be generated in the tWSe$_2$ and, simultaneously, nanoscale monolayer WSe$_2$ islands can be observed around the pit, implying that the tip pulses can locally tear the tWSe$_2$ into nanoscale islands (Extended Data Figs. 3 and 4). The most pronounced observation in this work is that the stacking configuration between the two layers of WSe$_2$ is changed dramatically around the pit, shown as significant changes in the moiré structure (Fig. 1d and Extended Data Figs. 4-6). After the STM tip pulses, the structure of the tWSe$_2$ manifests two distinct regions separating by a sharp boundary, as shown in Fig. 1d. In region I (around the pit), the moiré pattern of the WSe$_2$ disappears, indicating a localized rotation between the two layers of WSe₂ to the energetically more favorable R (or H) stacking configuration (Fig. 1e). In region II, there is a pronounced stretching of the moiré pattern, especially near the boundary between the two regions. According to the period of the moiré pattern around the boundary, the twist angle of the tWSe₂ in the region II is estimated as $\theta \sim 3.8° \pm 0.3°$. It indicates that the boundary separates two regions of the WSe₂ with different twist angles, indicating a large local lattice rotation and strain along the boundary. Our atomic-resolved STM measurements and the corresponding fast Fourier transform (FFT) images, as shown in Figs.1e-h, further reveal that there is a large lattice rotation, $\sim 3.0° \pm 0.6°$, of the topmost WSe₂ around the boundary. In tiny-angle twisted bilayer graphene (TBG), structure reconstruction leads to a local lattice rotation of about 1.0°, which can dramatically change electronic properties of the TBG[24-26]. In our experiment, the local lattice rotation along the 1D boundary is several times larger than that in structural-reconstructed tiny-angle TBG, implying that electronic properties of the 1D boundary could be quite different from that of its two adjacent regions.

Our STS measurement indicates that the electronic properties of the 1D boundary is changed dramatically by the large local lattice rotation: sharp peaks at negative energies are observed in the d$I$/d$V$ spectra recorded along the boundary, as shown in Figs. 2a-c (see Extended Data Fig. 7 for additional experimental data and similar results are obtained on different boundaries). In the tTMDs, such a phenomenon is a characteristic feature of the charging and discharging events of the local moiré sites with partially-filled flat bands[16,17]. The strongly localized correlated electrons in the partially-filled flat bands of the tTMDs not only lead to new correlated electronic phases[1-8], but also enable us to directly image the charge state of the local moiré sites by using STM[16,17]. When the moiré sites on the boundary gain or lose an electron, the Coulomb potential changes, altering the electron tunneling rate between the STM tip and the sample. This, in turn, leads to the occurrence of charging and discharging events, resulting in abrupt variations in the STM tunneling current, i.e., the sharp peaks in the STS spectra. Similar charging behavior has also been observed by STM in other nanoscale systems with strongly localized states[27-29]. As schematically shown in Fig. 2d, this characteristic phenomenon emerges as a consequence of partial filling of the flat bands in the valence band, where positive charges are present. Upon the application of a negative bias voltage, positive charges accumulate at the STM tip and repel neighboring positive charges of the moiré sites on the 1D boundary through local Coulomb blockade effects. As the positive charges accumulate to a critical level, a charging process occurs at the moiré site, resulting in the observation of a charging response in the negative energy range. The charging process also can be described by the tip-induced bending of the flat bands of local moiré sites, as illustrated subsequently.

According to our experiment, as shown in Figs. 2a-2c, the critical energy of the charging peaks sensitivity relies on the relative positions of the STM tip and the moiré site, regardless of the vertical or lateral direction. Figure 2c displays representative height-dependent d$I$/d$V$ spectra recorded at one of the moiré sites (see Extended Data Fig. 8 for more experimental data). As the STM tip approaches the moiré site (i.e., with increasing the tunneling current), the sharp peak in the spectrum shifts towards higher energies away from the Fermi level. The underlying reason for this phenomenon stems from the intimate relationship between the charging effect and the tip-induced band bending: the charging process corresponds to the tip-induced bending of the flat bands of local moiré sites, enabling it to cross the Fermi level and resonate with it. There are three factors that affect the band bending[16, 30-33]: the work function difference between the tip and tWSe$_2$ ($\Delta\Phi_{tip-tWSe_2}$), the tunneling sample bias voltage ($V_b$), and the back-gate voltage ($V_g$). Similar as previous studies[16, 30-33], the $\Delta\Phi_{tip-tWSe_2}$ elevates the moiré band beneath the tip to *p*-doping at zero sample bias in our work. A large positive bias voltage shifts the local potential distribution to deeper *p* doping and a large negative bias voltage shifts the local potential distribution to *n* doping.

Consequently, the unfilled moiré band of the local moiré site beneath the STM tip crosses the Fermi level under a negative bias, leading to the appearance of a charging peak. In the case of the STM tip approaching the sample in the vertical direction, the increased $\Delta\Phi_{tip-tWSe_2}$ results in pushing the moiré band further away from the Fermi level, necessitating a higher negative bias voltage on the tip for the charging process (Extended Data Fig. 9). In the lateral direction, when the STM tip is away from a specific moiré site, the tip-induced downward band bending at the moiré site is weaker. Consequently, a higher negative energy is required to facilitate the downward bending of the moiré band to cross the Fermi level. In the vicinity of the moiré site, the energy at which the charging peak appears is higher, as shown in Figs. 2a and 2b, which is expected to result in the formation of a ring centered around the moiré site.

To further explore charging ring structure around the moiré sites along the boundary, we carry out spatially dependent d$I$/d$V$ mapping. In the d$I$/d$V$ maps obtained at relatively high negative bias voltages, clear charging events were observed, characterized by prominent ring structures around the moiré sites along the boundaries, as shown in Figs. 3a and 3c (see Supplemental Material movies for more experimental results). As expected, the diameter of the charging rings depend sensitively on the bias voltage and the rings will even shrink to individual point at critical negative bias voltages. Interestingly, the individual charging point coincide with the moiré site near the boundary, indicating a strong correlation between the physical origin of the charging behavior and the moiré flat bands. However, our experiments reveal that the energy at which the charging peak appears depends sensitively on the moiré site. Figures 3b and 3d summarize the critical energies that the charging peak appears on the moiré sites along two different boundaries. Obviously, they are quite different among different moiré sites. In our experiment, the subtle difference of local strain and local lattice rotations on the moiré sites along the boundaries will lead to slight differences in the filling between different moiré sites, resulting in differences in the energy required for the moiré band to bend to the Fermi level (i.e., the different energies at which the charging peak appears). Previous studies have demonstrated explicitly that the tTMD moiré heterostructures provide a platform for simulating strongly correlated physics in the 2D extended Hubbard model[1-4,15], and the different energies at which the charging peak appear are observed in the STM measurements and they are described by different Hubbard on-site energies[16].

To further investigate the influence of filling on the charging effect, we applied a back gate to control the carrier density within the heterostructure. Figures 4a and 4b show two representative gate-dependent d$I$/d$V$ spectra obtained at two different sites, labelled as site 1 and site 2 respectively (see Extended Data Figs. 10 and 11 for details and see Extended Data Fig. 12 for more experimental results). The site 1 is near the center of one moiré site, therefore, we only detect the charging/discharging peak of one moiré site. The site 2 is a high-symmetry point with almost equidistant to three neighbouring moiré sites. As a consequence, we can observe the charging/discharging peaks of the three moiré sites. Obviously, in both cases, the energies of the charging/discharging peaks depend sensitively on the back-gate voltage. With increasing the back-gate voltage from $V_g$ = -20 V to $V_g$ = 16 V, one of the flat moiré bands is tuned from positive energy to approach the Fermi level. Therefore, a smaller negative bias voltage on the tip is enough to bend the flat moiré band to the Fermi level for the charging process, as observed in Figs. 4a and 4b. Such a behavior confirms that the charging behavior is determined by the filling of the flat moiré band. Our d$I$/d$V$ mappings at a back-gate voltage of 15 V, as summarized in Fig. 4c, further demonstrate the above result (see Supplemental Material movies for more experimental results). There are two distinct charging behaviors of different moiré sites along the boundary. In case 1, the diameter of the charging ring for some moiré sites increases by further increasing the negative bias voltages, as observed at 0 V back-gate voltage in Fig. 3. As an example, the charging ring observed at -410 mV is larger than that observed at -380 mV (Fig. 4c). In case 2, the opposite behavior is observed and the diameter of the charging ring for the other moiré sites decreases by further increasing the negative bias voltages. For example, the diameter of the charging ring observed at -77 mV is larger than that observed at -135 mV. All these phenomena can be understood self-consistently with the flat moiré band filling picture. In the case 1, the moiré flat band for some moiré sites is slightly above the Fermi level in the absence of tip-induced bending. As the negative bias voltage increases, the moiré band of these moiré sites bends towards the Fermi level, leading to the appearance of charging peaks. With further increasing the negative bias voltage, the degree of band bending increases, resulting in an increase in the diameter of the charging ring, as schematically shown in Fig. 4c. In the case 2, the moiré flat band for the other moiré sites is completely filled in the absence of tip-induced bending. Then, the tip-induced band bending allows us to detect the charging peaks at a small bias voltage. As the negative bias voltage increases, the band gradually bends downward, reducing the degree of upward band bending and causing the diameter of the charging ring to decrease until it disappears, as schematically shown in Fig. 4c. The emergence of two distinct charging behaviors of different moiré sites further confirms the different fillings of the moiré sites along the boundary, as shown in Figs. 2 and 3. The different fillings, i.e., the subtle energy difference between the flat moiré band and the Fermi level, in different moiré sites in turn results in different band bending effects under various bias voltage and back-gate voltage.

In conclusion, we present a facile and general approach to implement 1D moiré chains with partially-filled flat bands within 2D TMD systems. This methodology establishes a pioneering moiré superlattice platform for simulation of strongly correlated physics that is governed by the 1D Hubbard model. The reported results pave the way for expanding the 2D correlated moiré physics to 1D or even fractional dimensional, offering unprecedented opportunities for studying and comprehending the intricacies of strongly correlated physics within these systems.

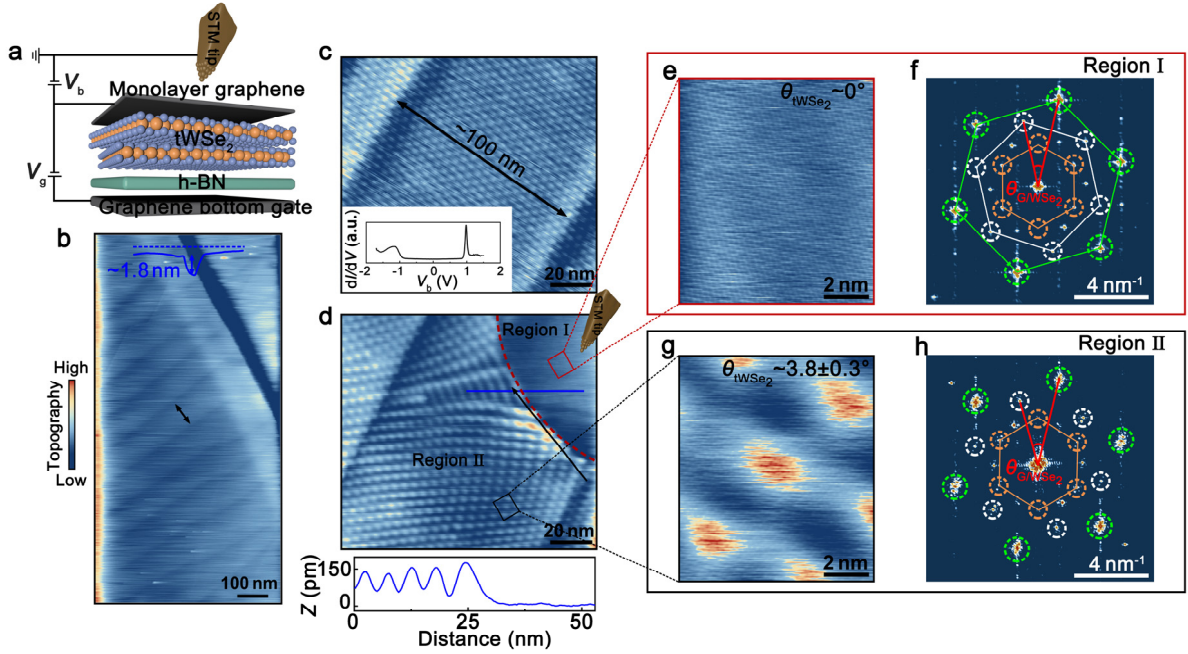

**Fig. 1 | In situ creating 1D moiré chain in graphene/tWSe₂ heterostructure. a,** Sketch of graphene/tWSe$_2$ heterostructure device used for STM study. **b,** A large-scale STM topography image ($V_b$ = 1200 mV and $I$ = 200 pA) measured on the sample surface, revealing a quasi-1D superlattice with an approximate period of 100 nm. This superstructure is in the hBN substrate, possibly generating by its 1D crack (see inset for profile line of the 1D crack). **c,** A representative zoom-in STM topographic image ($V_b$ = -400 mV and $I$ = 100 pA) reveals a moiré superlattice morphology with a period of about 4.4 nm. Inset, a typical d$I$/d$V$ spectrum, showing two peaks around the valence and conduction band edges, respectively. **d,** Top panel: STM topographic image ($V_b$ = -400 mV and $I$ = 300 pA) after applying a tip voltage pulse. The moiré structure observed in the image exhibits two distinct regions. In region I, the moiré pattern completely vanishes. In region II, a conspicuous stretching of the moiré pattern is observed. Bottom panel: A height profile acquired by measuring along the blue line in the top panel, spanning across the two regions. **e** and **g,** Atomic-resolved STM images in the region I and region II, respectively. **f** and **h,** The FFT images obtained from the STM images of region I and region II, respectively. The green and white circles show reciprocal lattices of graphene and WSe$_2$, respectively. The orange circles

show reciprocal moiré lattices of the graphene/WSe$_2$ heterostructure. The twist angles between WSe$_2$ and graphene can be obtained from the FFT, ~27.1° in the region I and ~30.1° in the region II respectively. The relative lattice rotation of WSe$_2$ along the boundary can be inferred as $\theta \sim 3.0 \pm 0.6°$, consistent with that calculated through the WSe$_2$ moiré patterns of the two adjacent regions.

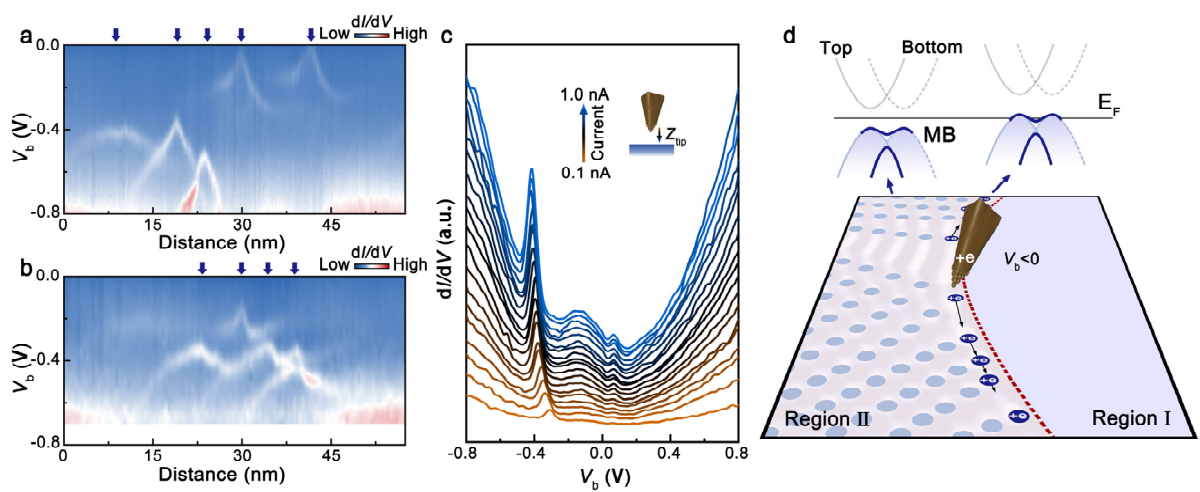

**Fig. 2 | Local charging effect of moiré sites along the 1D boundary. a** and **b,** The d$I$/d$V$ spectroscopic maps versus the spatial position along the direction of black solid lines (a) in Fig. 1d and (b) in Extended Data Fig. 6, respectively. **c,** The d$I$/d$V$ spectra measured at different tip-sample distances $Z_{tip}$. The tip height decreases by increasing the tunneling current $I$ with a fixed voltage bias. When the tip is approaching the sample (as the tunneling current increases), the charging peak moves away from the Fermi level. **d,** Sketch of the tip-induced charging effect at the moiré sites along the 1D boundary. Under significant strain and local lattice rotations, the valence moiré band (MB) at the boundary moiré sites is tuned to approach the Fermi level. The STM tip serves as a local top gate, and when the applied bias voltage exceeds a certain threshold, the flat band crosses the Fermi level, leading to a charging response.

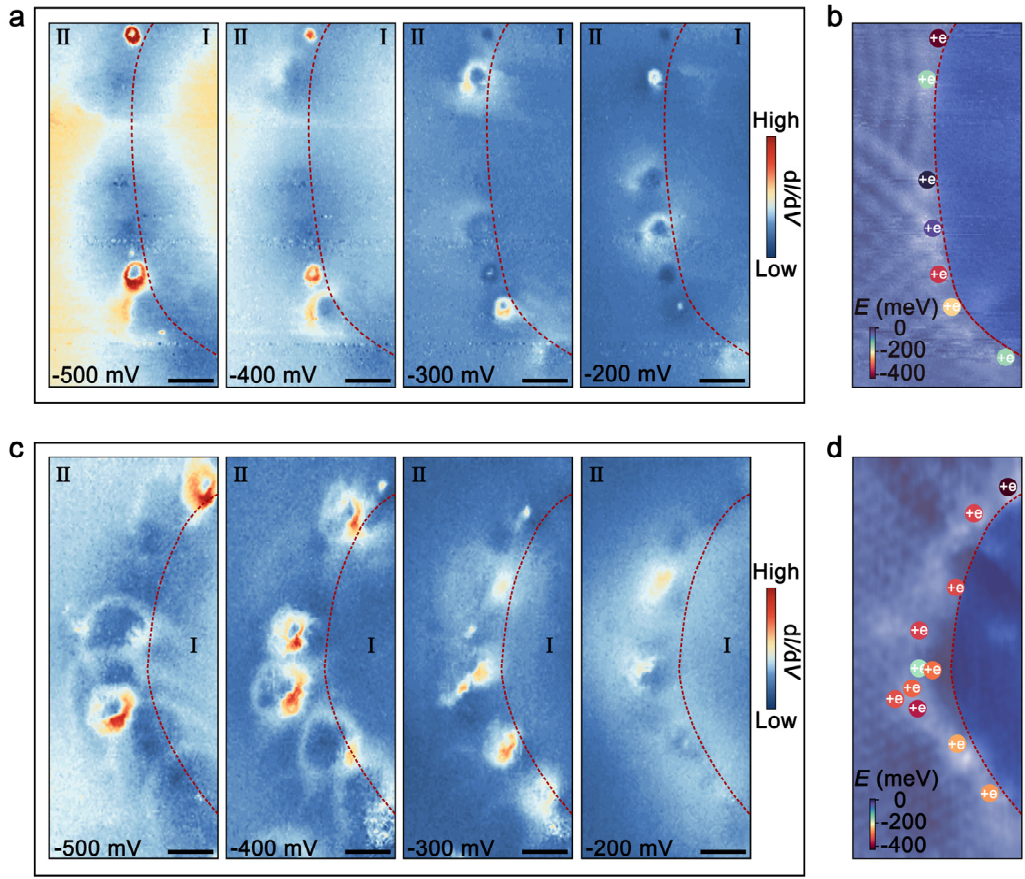

**Fig. 3 | Evolution of moiré charging rings with changing $V_b$. a** and **c,** Spatial dependent d$I$/d$V$ mappings on two different boundary regions. The scale bar is 10 nm. **b** and **d**, A statistical summary showing the critical energy at which the charging peak appears for each moiré site in both regions, respectively. The different colors represent different energy values.

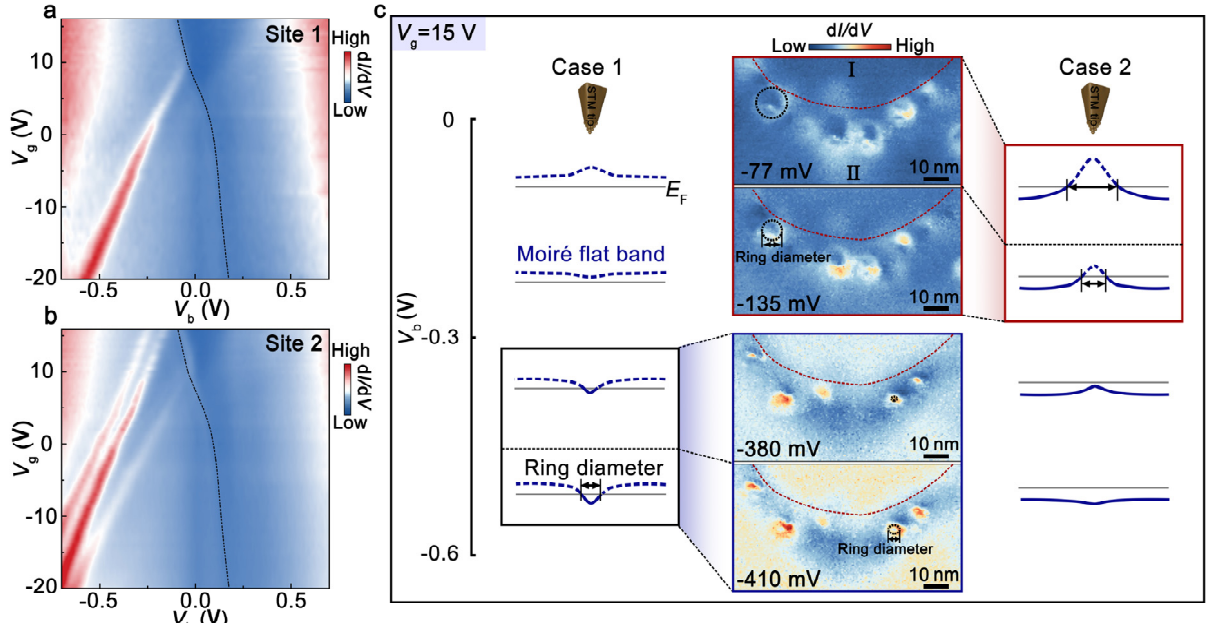

**Fig. 4 | The impact of electron fillings on the charging effect of moiré sites. a** and **b,** Gate-dependent d$I$/d$V$ spectra measured at two different moiré sites. The black dashed lines indicate the energy of the graphene Dirac point extracted by measuring the d$I$/d$V$ spectra far from the 1D moiré sites (~ 30 nm) at various back-gate voltages. **c,** Middle panel: Spatially-dependent d$I$/d$V$ mappings at different energies with a back-gate voltage of 15V. Left and right panels: The schematic diagram of the energy band filling as a function of bias voltage for two different cases. In Case 1, the diameter of the charging ring observed at high energy (-410 meV) is greater than that observed at low energy (-380 meV). In Case 2, the opposite behavior is observed and the diameter of the charging ring observed at low energy (-77 meV) is greater than that observed at high energy (-135 meV). The distinct behaviors arise from the different filling of the flat band in these moiré sites.

# Methods

**Sample fabrication.** The graphene/tWSe$_2$ heterostructure was fabricated using the micro-mechanical stacking technique[17-20] based on the transfer platform from Shanghai Onway Technology Co., Ltd.. By utilizing a polydimethylsiloxane (PDMS) film stamp, the process of sequentially picking up all exfoliated two-dimensional material flakes on the PDMS film is performed in the following order: the bottom hBN layer, the first layer of WSe$_2$, the second layer of WSe$_2$, and then monolayer graphene. The two WSe$_2$ layer achieves relative rotation of approximately 4° based on straight edge alignment. Subsequently, the entire heterostructure is flipped using another PDMS film, and the original PDMS film is removed. The heterostructure is then placed on a SiO$_2$/Si substrate, which is pre-patterned with few-layered graphene or gold electrodes. Finally, electrical contact is established by connecting the electrodes to the sample surface using a graphite thin film.

**STM measurements.** STM/STS measurements were performed in low-temperature (77 K) and ultrahigh-vacuum (~10$^{-10}$ Torr) scanning probe microscopes (USM-1400) from UNISOKU. The tips were obtained by chemical etching from a tungsten wire. The differential conductance (d$I$/d$V$) measurements were taken by a standard lock-in technique with an ac bias modulation of 5 mV and 793 Hz signal added to the tunneling bias.

## Acknowledgments

This work was supported by the National Key R and D Program of China (Grant Nos. 2021YFA1401900, 2021YFA1400100), National Natural Science Foundation of China (Grant Nos. 12141401, 11974050), and "the Fundamental Research Funds for the Central Universities"

## Author contributions

Y. N. R. and L. H. designed the experiment. Y. N. R. synthesized the samples, performed the STM experiments. K.W. and T.T. provided the hexagonal boron nitride crystals. Y. N. R. and L. H. analyzed the data and wrote the paper. All authors participated in the data discussion.

## Data availability statement

All data supporting the findings of this study are available from the corresponding author upon reasonable request.